\documentclass[]{spie}  

\usepackage{amsmath,amsfonts,amssymb}
\usepackage{graphicx}
\usepackage[colorlinks=true, allcolors=blue]{hyperref}

\title{Design and characterization of a ground-based absolute polarization calibrator for use with polarization sensitive CMB experiments}

\author[]{M. F. Navaroli}
\author[]{G. P. Teply}
\author[]{K. D. Crowley}
\author[]{J. P. Kaufman}
\author[]{N. B. Galitzki}
\author[]{K. S. Arnold}
\author[]{B. G. Keating}
\affil[]{The University Of California, San Diego}

\authorinfo{}

\begin{document} 
\maketitle

\begin{abstract}
We present the design and characterization of a ground-based absolute polarization angle calibrator accurate to better than 0.1$^\circ$ for use with polarization sensitive cosmic microwave background (CMB) experiments. The calibrator's accuracy requirement is driven by the need to reduce upper limits on cosmic polarization rotation, which is expected to be zero in a large class of cosmological models. Cosmic polarization effects such as cosmic birefringence and primordial magnetic fields can generate spurious B-modes that result in non-zero CMB TB and EB correlations that are degenerate with a misalignment of detector orientation. Common polarized astrophysical sources used for absolute polarization angle calibration have not been characterized to better than 0.5$^\circ$. Higher accuracy can be achieved through self-calibration methods, however these are subject to astrophysical foreground contamination and inherently assume the absence of effects like cosmic polarization rotation. The deficiencies in these two calibration methods highlight the need for a well characterized polarized source. The calibrator we present utilizes a 76 GHz Gunn oscillator coupled to a frequency doubler, pyramidal horn antenna, and co-rotating wire-grid polarizer. A blackened optics tube was built around the source in order to mitigate stray reflections, attached to a motor-driven rotation stage, then mounted in a blackened aluminum enclosure for further reflection mitigation and robust weatherproofing. We use an accurate bubble level in combination with four precision-grade aluminum planes located within the enclosure to calibrate the source's linear polarization plane with respect to the local gravity vector to better than the 0.1$^\circ$ goal. In 2017 the calibrator was deployed for an engineering test run on the POLARBEAR CMB experiment located in Chile's Atacama Desert and is being upgraded for calibration of the POLARBEAR-2b receiver in 2018. In the following work we present a detailed overview of the calibrator design, systematic control, characterization, deployment, and plans for future CMB experiment absolute polarization calibration.
\end{abstract}


\section{INTRODUCTION}
\label{sec:intro}  

The cosmic microwave background (CMB) is the oldest observable electromagnetic radiation in the universe and offers a detailed snapshot the universe at the end of the epoch of recombination 378,000 years after the Big Bang, providing an invaluable test bed for fundamental physics. The CMB was partially polarized at the surface of last scattering and its polarization can be separated into even-parity and odd-parity modes called E-modes and B-modes~\cite{Seljak1996,Kamionkowski1997,Kamionkowski1997a}. Density perturbations at the the time of last scattering generated E-mode polarization in the CMB~\cite{Ade2015d} while effects with odd-parity components such as primordial gravitational waves~\cite{Kamionkowski1997,Kamionkowski1997a,Crittenden1993}, gravitational lensing~\cite{Ade2014,Lewis2006}, and cosmic polarization rotation (from either cosmic birefringence~\cite{Carroll1998,Lue1999} or primordial magnetic fields~\cite{Loeb1996,Subramanian1998,Seshadri2001}) generate both E-mode and B-mode polarization. The CMB is a near-perfect, isotropic, 2.73 K blackbody across the sky with fluctuations smaller than one part in ten thousand.~\cite{Mather1994a,Smoot1992} The temperature anisotropies, $\Delta_T(\theta,\phi)$, can be decomposed into spherical harmonics $Y_{\ell m}$ with coefficients $a_{\ell m}^T$ as
\begin{equation}
\label{eq:sh}
\Delta_T(\theta,\phi) = \sum\limits_{\ell=1}^{\infty} \sum\limits_{m=-\ell}^\ell a_{\ell m}^T Y_{\ell m}(\theta,\phi) \, .
\end{equation}
For Gaussian fluctuations, all of the statistical behavior of the spherical harmonic coefficients $a_{\ell m}$ is captured by their power spectrum $C_\ell$, defined as
\begin{equation}
\label{eq:cell}
\langle a_{\ell m}^{X} a^{*X'}_{\ell 'm'} \rangle = \delta_{\ell \ell '} \delta_{mm'} C^{XX'}_\ell \, ,
\end{equation} 
where $X$ and $X'$ are combinations of T, E, and B. The even-parity and odd-parity correlations $C_\ell^{TB}$ and $C_\ell^{EB}$ are expected to vanish according the standard model but can be non-zero if CMB detectors are misaligned or if the CMB polarization is rotated by effects such as cosmic polarization rotation.

\subsection{Cosmic polarization rotation and the CMB}

Linearly polarized electromagnetic radiation is said to undergo cosmic polarization rotation (CPR) by angle $\alpha$ if the radiation's polarization plane experiences a rotation as it traverses cosmological distances~\cite{Lue1999}. For instance, if our universe were inherently birefringent it could generate non-zero CPR by inducing a relative phase lag between the two polarization states of propagating radiation, causing the radiation's polarization plane to be rotated by some angle $\alpha$.~\cite{Carroll1998}. A non-zero measurement of this so-called ``cosmic birefringence" could be an indication of charge, parity, and time-reversal symmetry violating physics in our universe~\cite{Kaufman2016,Ade2015d}. To search for evidence of cosmic birefringence we must observe linearly polarized radiation that has traveled significant cosmological distances and has a known emitted polarization. Several observations of the polarized radio and UV synchrotron emission from galaxies and quasars have been made and have found potential cosmic birefringence rotation angles $\alpha$ to be weakly consistent with a small negative rotation~\cite{Carroll1998,Kamionkowski2010,Wardle1997,Alighieri2010}. Another phenomenon that could generate non-zero CPR occurs when primordial magnetic fields (PMFs) embedded in the photon-baryon plasma at the surface of last scattering Faraday rotate the polarization planes of CMB photons by an angle $\alpha$ along line of sight $\boldsymbol{n}$ given by \cite{Ade2015d,Loeb1996,Harari1997}
\begin{equation}
\label{eq:pmfalpha}
\alpha(\boldsymbol{n}) = \frac{3c^2}{16 \pi^2 e} \nu^{-2} \int \dot\tau \boldsymbol{B} \cdot d \ell \, ,
\end{equation}
where $e$ is the electron charge, $\nu$ is the observed frequency of the radiation, $\dot \tau$ is the differential optical depth, $\boldsymbol{B}$ is the co-moving magnetic field, and $d \ell$ is the co-moving length element along the photon trajectory.

The polarized CMB is perhaps the most promising candidate to measure non-zero CPR effects such as cosmic birefringence and PMFs. As CMB photons have traveled the largest cosmological distance of any observable radiation in the universe they would have undergone the largest polarization angle rotation $\alpha$ if our universe were birefringent. Their presence at the surface of last scattering makes CMB photons an ideal test subject to measure Faraday rotation by PMFs in the early universe. While cosmic birefringence and PMFs both generate polarization rotation, the two effects can be differentiated by noting the strong frequency dependence in Eq.~\ref{eq:pmfalpha} and observing the CMB in multiple frequency bands. 

\begin{figure} [ht]
	\begin{center}
		\begin{tabular}{c} 
			\includegraphics[height=7cm]{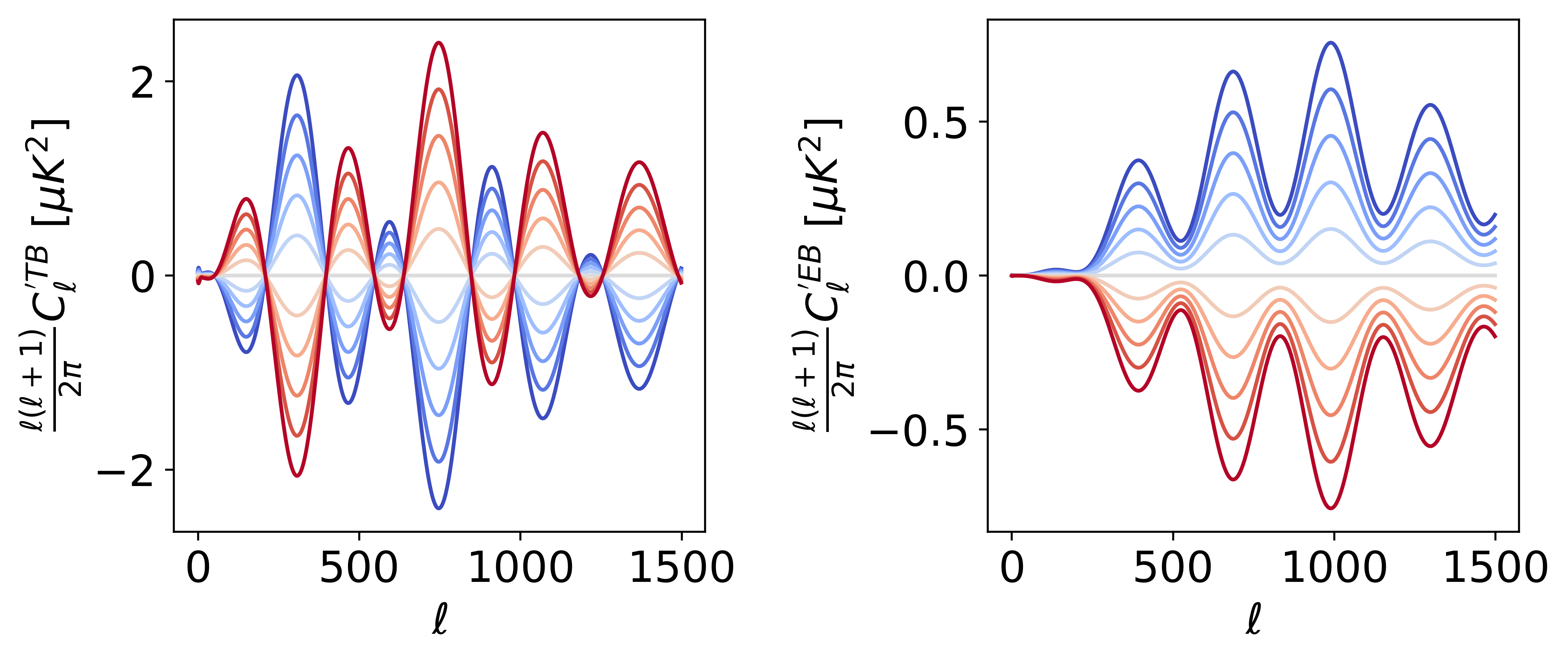}
		\end{tabular}
	\end{center}
	\caption[example] 
	{ \label{fig:cleb_cltb} 
		$C_\ell^{'TB}$ (left) and $C_\ell^{'EB}$ (right) power spectra after applying a -2.5$^\circ$ (darkest blue) to +2.5$^\circ$ (darkest red) polarization rotation to the primordial $C_\ell^{TB}$ and $C_\ell^{EB}$ power spectra in $0.5^\circ$ steps. Note that $C_\ell^{'TB}$ and $C_\ell^{'EB}$ are both zero for $\alpha = 0$ as per the standard model.}
\end{figure}

In the standard cosmological model the CMB power spectra $C_\ell^{TB}$ and $C_\ell^{EB}$ are expected to vanish due to symmetry\cite{Kamionkowski1997}. If an effect, whether astrophysical or systematic, rotates the CMB polarization it mixes E-modes into B-modes. These spurious B-modes generate non-zero TB and EB correlations that leaks $C_\ell^{TE}$ to $C_\ell^{TB}$ and $C_\ell^{EE}$ to $C_\ell^{BB}$ as\,~\cite{Lue1999,Keating2013}
\begin{equation}
\label{eq:clmix}
\begin{split}
&C_\ell^{'TT} = C_\ell^{TT} \\
&C_\ell^{'TE} = C_\ell^{TE}\cos{(2\alpha)} \\
&C_\ell^{'EE} = C_\ell^{EE}\cos^2{(2\alpha)} + C_\ell^{BB}\sin^2{(2\alpha)} \\
&C_\ell^{'BB} = C_\ell^{EE}\sin^2{(2\alpha)} + C_\ell^{BB}\cos^2{(2\alpha)} \\
&C_\ell^{'TB} = -\,C_\ell^{TE}\sin{(2\alpha)} \\
&C_\ell^{'EB} = -\,\frac{1}{2}(C_\ell^{EE} - C_\ell^{BB})\sin{(4\alpha)} \, ,
\end{split}
\end{equation} 
where the $C_\ell^{'XX}$ are the observed angular power spectra, the $C_\ell^{XX}$ are the theoretical non-rotated angular power spectra, and $\alpha$ is the applied rotation angle. An example of the rotated $C_\ell^{'TB}$ and $C_\ell^{'TE}$ power spectra for several values of $\alpha$ are shown in Fig.~\ref{fig:cleb_cltb}. These non-zero TB and EB correlations generated by cosmological sources are degenerate with a systematic misalignment of instrument detector orientation~\cite{DeBernardis2016,Keating2013,Nati2017}, which is the only instrumental systematic error that is capable of producing TB and EB correlations with a common rotation angle~\cite{Kaufman2016,Miller2009a}. Therefore, in order to measure CPR effects such as cosmic birefringence and PMFs, absolute detector polarization orientation must be strictly calibrated. 

\subsection{State of the art and current calibration limitations}
Both current and future generation CMB polarization experiments are in a promising position to measure potential effects from CPR. Many experiments are either already observing in multiple frequency bands or plan to in order to separate CMB signal from astrophysical foregrounds. The most commonly used polarized astrophysical source for absolute detector polarization angle calibration of CMB experiments is Taurus A (henceforth ``Tau A"), a supernova remnant also called the Crab Nebula. However, the polarization angle of Tau A has only been calibrated to $\pm 0.5^\circ$~\cite{Aumont2010b}. Other polarized astrophysical sources such as Centaurus A have been characterized to comparable levels~\cite{Zemcov2010}. In addition to using polarized astrophysical sources to calibrate absolute polarization of detectors, CMB experiments can self-calibrate by forcing $C_\ell^{TB}$ and $C_\ell^{EB}$ to be zero as per the standard cosmological model, though doing so assumes effects like CPR and PMFs to be zero~\cite{Keating2013}. Several CMB experiments have used both astrophysical and artificial polarized sources to constrain TB and EB correlations, and therefore also the rotation angle $\alpha$~\cite{Komatsu2011,Hinshaw2013,Pagano2009,Wu2009,Li2015,Collaboration2012,Kaufman2014,Ade2014e,Ade2014a,Koopman2016a,Naess2014}. However, studies suggest that a joint analysis of the data sets yield no detection of a non-zero value for $\alpha$ as the large systematic uncertainties (of order 0.5$^\circ$ to 1.5$^\circ$) dominate the statistical uncertainties~\cite{Galaverni2015b,Li2015,Xia2010}.

The deficiencies in calibrating absolute detector angles via astrophysical and artificial sources or via self-calibration are clear. Polarized astrophysical sources have not been calibrated well enough to measure values of $\alpha < 0.5^\circ$, systematic uncertainties in detector calibration angle dominate statistical uncertainties, and self-calibration assumes polarization rotation $\alpha$ from cosmological origin to be zero. In order to calibrate absolute detector polarization angle to better than $0.5^\circ$ to potentially detect CPR, a well characterized polarized source must be developed. In the following, we present the design and characterization of a ground-based CMB experiment absolute polarization calibrator that utilizes tight control over systematics in order to achieve calibration angles of $\alpha < 0.1^\circ$. Section~\ref{sec:instrument} details the design of this calibrator while Section~\ref{sec:analysis} presents the data analysis procedure and laboratory characterization results. Section~\ref{sec:future} describes plans for the calibrator's improvement and Section~\ref{sec:conclusions} concludes with remarks on plans for the calibrator to be used with future CMB experiments.

\section{GROUND-BASED POLARIZATION CALIBRATOR}
\label{sec:instrument}

Several ground-based absolute polarization angle calibrators have been designed and tested with current CMB experiments. Common calibration sources include polarized thermal sources, dielectric sheets, and polarizing wire-grids~\cite{Masi2006,Wu2009,Kaufman2014,Ade2014a,Kusaka2018}. In this section we focus on the methodology and design of a rotating polarized microwave source with a co-rotating, polarizing wire grid (henceforth ``the calibrator") that was deployed on the POLARBEAR telescope in 2017 for an initial engineering test run to prove calibrator-telescope coupling and field operation procedure.

\subsection{CALIBRATOR DESIGN AND HARDWARE}

\subsubsection{Rotating chopped polarized source}
\begin{figure} [ht]
	\begin{center}
		\begin{tabular}{c} 
			\includegraphics[height=6cm]{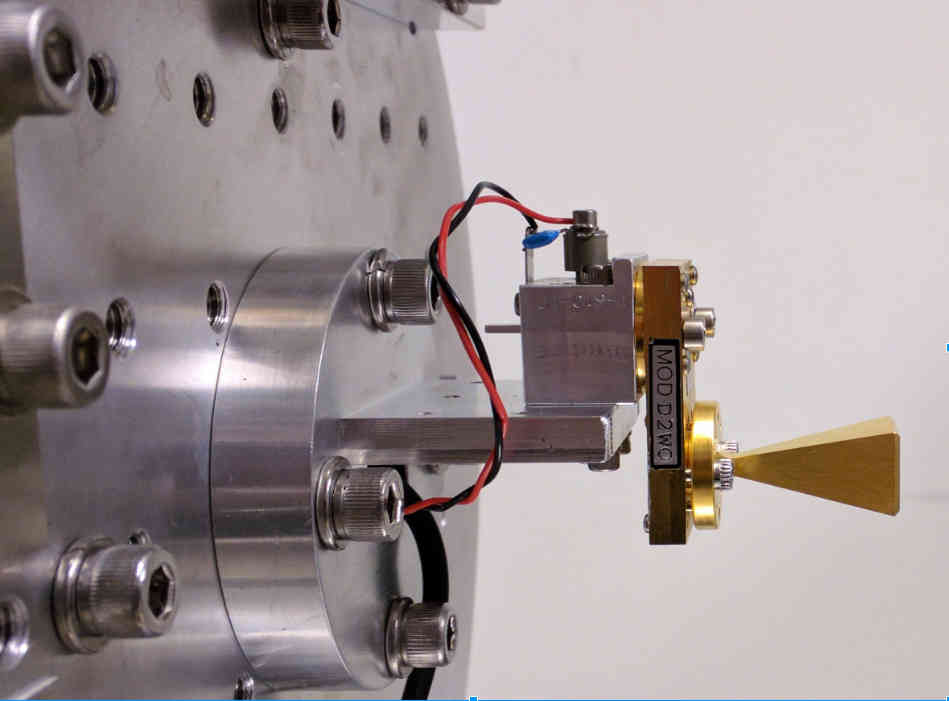}
			\includegraphics[height=6cm]{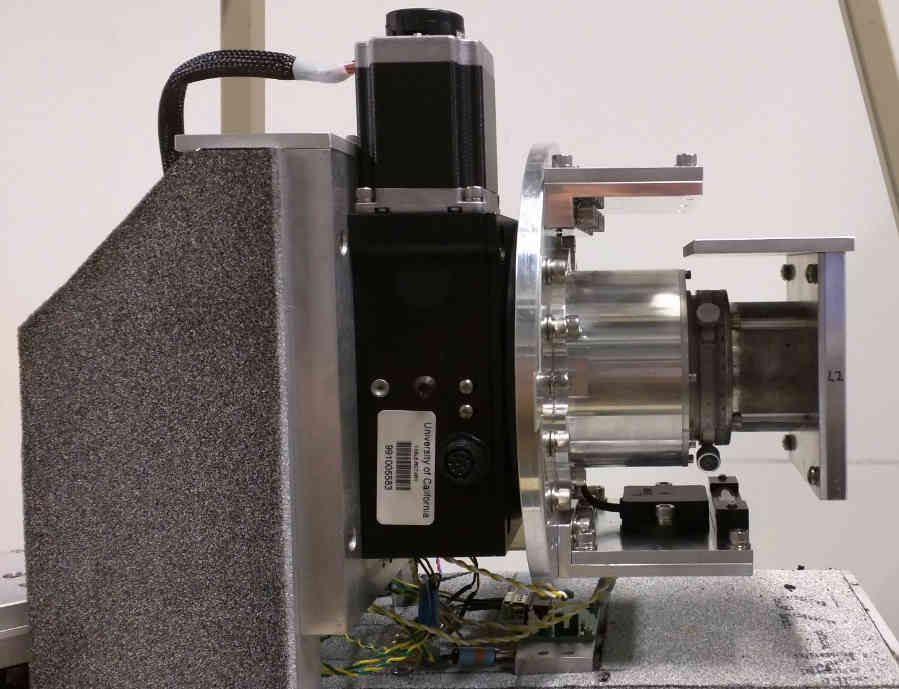}
		\end{tabular}
	\end{center}
	\caption[example] 
	{ \label{fig:sourcecross} 
		\textit{Left}: The Gunn oscillator + doubler + pyramidal horn circuit mounted on the rotation stage. \textit{Right}: The co-rotating blackened aluminum cylinder with adjustable co-rotating polarizing wire grid mounted around the Gunn oscillator circuit.}
\end{figure}

As nearly every CMB experiment includes a wide observation band centered near 150 GHz, the calibrator utilizes a 76 GHz Gunn oscillator coupled to a frequency doubler. The Gunn oscillator is further coupled to a Sage Millimeter\footnote{Sage Millimeter, Inc. Torrance, CA. www.sagemillimeter.com} WR-06 pyramidal horn antenna in order to transmit a linearly polarized signal. The power output of the oscillator was measured to be 0.75 mW, integrated across the beam described in section~\ref{sec:labtests}, by connecting the waveguide output of the frequency doubler to the waveguide input of a microwave power meter. The Gunn oscillator with doubler and horn antenna are mounted on a motorized rotation stage with 0.001$^\circ$ accuracy encoder readout in order to rotate the calibrator's linearly polarized signal by precisely known angles. A co-rotating aluminum cylinder coated with Berkeley Black~\cite{Persky1999} is mounted around the Gunn oscillator to mitigate stray reflections. To further ensure linearly polarization purity of the source, a co-rotating polarizing wire grid is mounted to the end of the blackened aluminum cylinder to a lockable 0.002$^\circ$ accuracy stage to allow fine adjustment of the wire grid polarization axis with respect to the horn antenna. The Gunn oscillator circuit, aluminum cylinder with wire grid, and rotation stage system (henceforth ``the source") are shown in Fig.~\ref{fig:sourcecross}. In order to differentiate the linearly polarized source signal from the ground and other systematics, a chopper wheel is installed between the polarizing wire grid and the front of the enclosure. Two photodiodes placed with respect to the chopper blades at 0$^\circ$ and 45$^\circ$, which correspond to 0$^\circ$ and 90$^\circ$ in phase space due to the chopper consisting of two blades, act as quadrature encoders.

\subsubsection{Enclosure and pointing}
The calibrator's motorized rotation stage is mounted inside a rectangular aluminum enclosure in order to provide ruggedness for handling and protection from the harsh environment in Chile. All inner-facing aluminum is coated with Eccosorb\footnote{Emerson \& Cuming Microwave Products, a unit of Laird Technologies. Chesterfield, MO. www.eccosorb.com} AN-72 to mitigate potential reflections. A circular window is cut out of the front-facing side of the enclosure to allow propagation of the source signal with a diameter set to be 1.5 times larger than the beam full width at half maximum (FWHM) to reduce any diffraction effects that might interfere with the beam. Two weather-tight materials are used for the window: high density polyethylene foam, effectively transparent at 150 GHz, and Eccosorb MF-110 plastic, measured to attenuate a 150 GHz signal by 15.4 $\pm$ 0.6 dB per centimeter thickness. Four modular MF-110 windows with thicknesses 0.635 cm each provided variable attenuation up to 39 dB.

\begin{figure} [ht]
	\begin{center}
		\begin{tabular}{c} 
			\includegraphics[height=7cm]{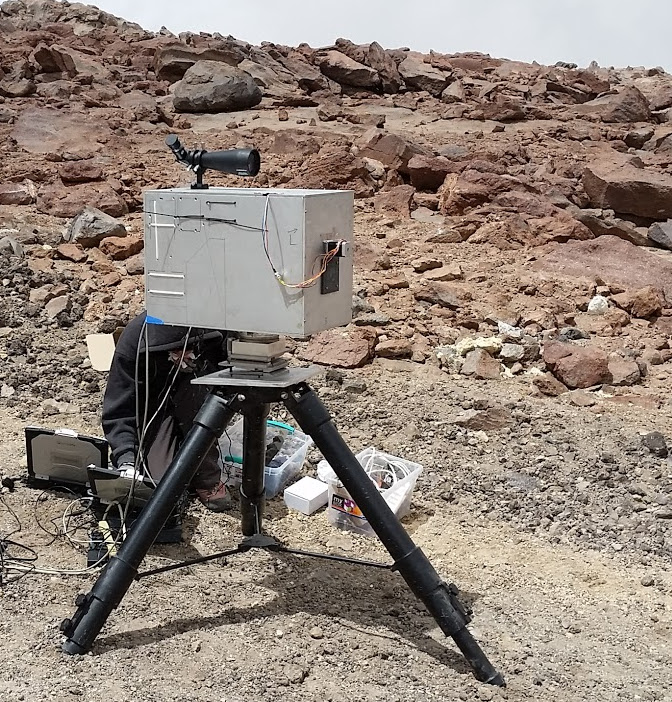}
			\includegraphics[height=7cm]{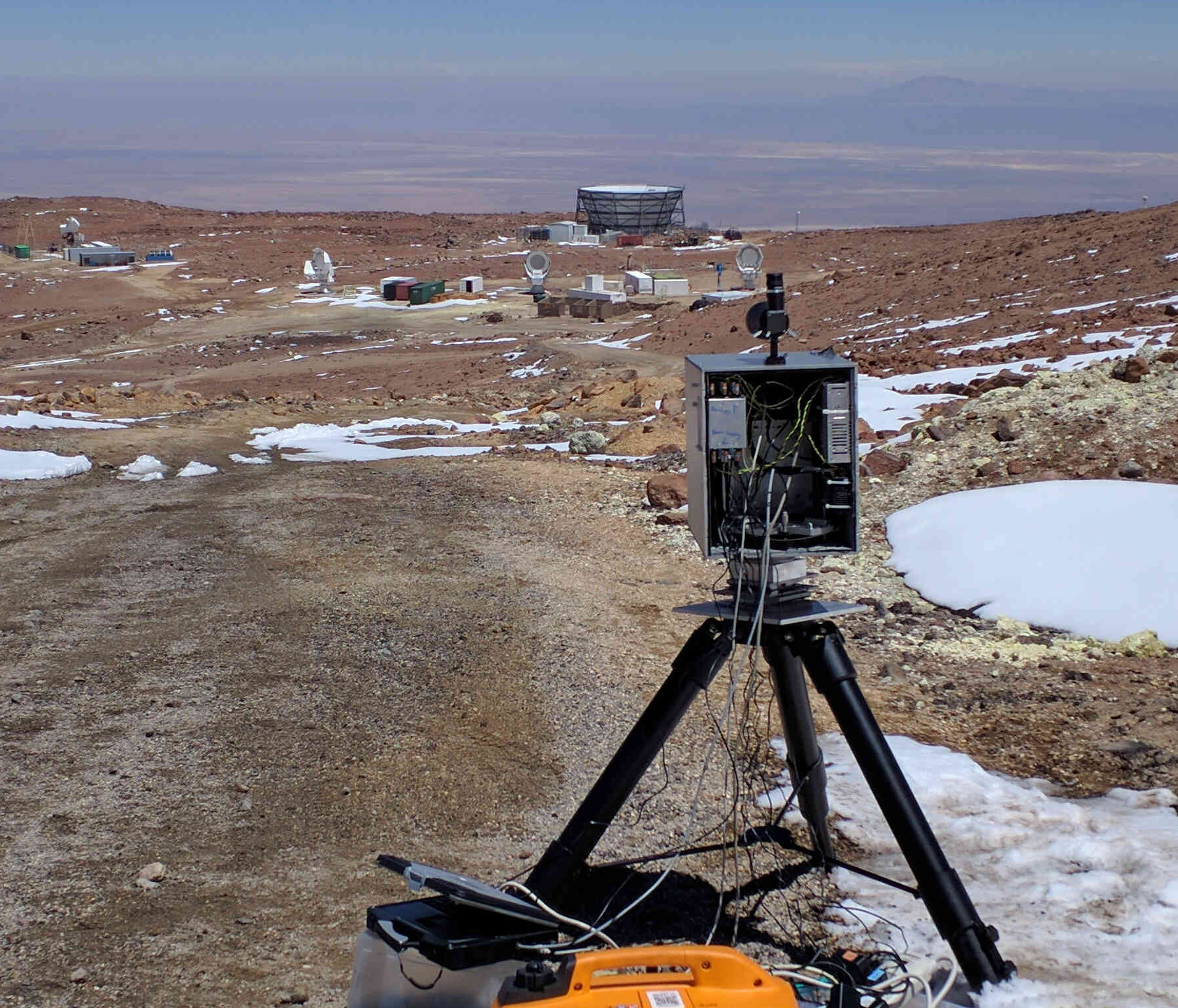}
		\end{tabular}
	\end{center}
	\caption[example] 
	{ \label{fig:fieldsetup} 
		\textit{Left}: The full polarization calibrator setup at the POLARBEAR site in the Atacama Desert, Chile. \textit{Right}: The calibrator pointed towards the POLARBEAR telescope.}
\end{figure}

The enclosure is mounted on an adjustable azimuthal yaw stage which, in combination with a finder scope mounted on top of the enclosure, can be adjusted for rough pointing towards the telescope. Underneath the yaw stage are two perpendicular one-axis, 0.004$^\circ$ accuracy tilt stages and a 0.001$^\circ$ accuracy yaw stage for output horizontal and vertical polarization angle adjustment. The enclosure and tilt and yaw stages are mounted to a 23 kg Meade\footnote{Meade Instruments. Irvine, CA. www.meade.com} Giant Field Tripod to provide a sturdy base. The Gunn oscillator power output varies with temperature by -0.4 dB/C$^\circ$ and can be stabilized with self-regulating heating pads placed on the outside of the calibrator enclosure, though the pads were not used in the engineering deployment described in section~\ref{sec:deployment}. A thermometer is placed near the Gunn oscillator and measured throughout a calibration in order to account for the varied output power. Images of the full calibrator setup while deployed on an initial engineering test run on the POLARBEAR telescope in the Atacama Desert, Chile in March 2017 are shown in Fig.~\ref{fig:fieldsetup}.

\subsection{Calibration to the local gravity vector}
\label{sec:calgrav}

\begin{figure} [ht]
	\begin{center}
		\begin{tabular}{c} 
			\includegraphics[height=7cm]{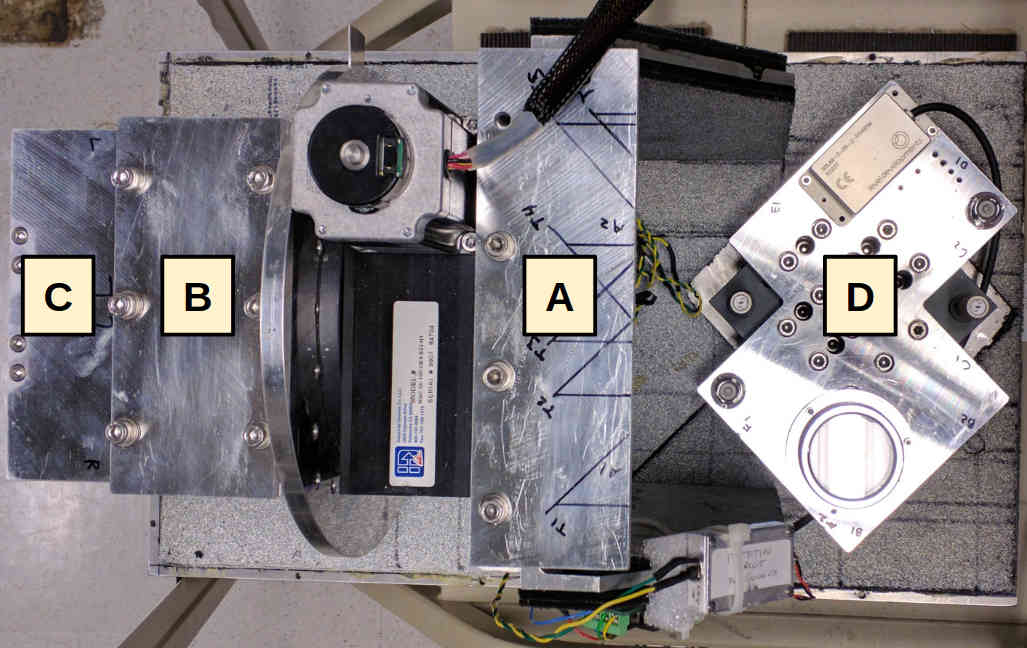}
		\end{tabular}
	\end{center}
	\caption[example] 
	{ \label{fig:planes} 
		The four precision-ground aluminum planes used for calibrating to the local gravity vector. Planes A, B, and C are fixed with respect to the enclosure, rotation stage, and co-rotating wire grid respectively. Plane D is mounted on a 0.001$^\circ$ accuracy two-axis tilt stage with micrometer dials (fixed with respect to the enclosure) to allow horizontal and vertical adjustment of the plane. Angle measurements of each combination of micrometer dial value were made using a digital tilt meter to determine the angle at each micrometer dial value.}
\end{figure}

We calibrate the initial polarization angle of the source to the local gravity vector to better than the 0.1$^\circ$ science target by using a 0.006$^\circ$ accuracy bubble level in combination with four precision-ground aluminum planes located within the enclosure. The calibrator is first pointed azimuthally toward the target using the yaw stage beneath the enclosure. Planes A, B, C, and D,  as labeled in Fig.~\ref{fig:planes}, are then leveled by adjusting the two perpendicular, one-axis tilt stages beneath the enclosure,  the rotation stage motor position, the polarizing grid's yaw stage, and the micrometer dials on the two-axis tilt stage respectively. The enclosure is then pointed downwards towards the target by using the vertical one-axis tilt stage until the target appears in the optical finder scope. Plane D is then re-leveled and the calibrator's output polarization plane is determined with respect to the local gravity vector by comparing the initial and final values of the two-axis tilt stage micrometer dials. A two-axis 0.01$^\circ$ accuracy digital tiltmeter is placed on plane D and continuously monitored throughout a measurement to account for any shifts in source calibration angle due to winds, vibrations, or other effects that might permanently shift the calibrator enclosure.

\subsection{Control software and data acquisition}

Both the chopper and rotation stage stepper motors are controlled via a Moxa\footnote{Moxa. Chesterfield, MO. www.moxa.com} NPort 5450 serial device server. The chopper stepper motor is commanded to a single rotation frequency throughout the duration of a measurement while the rotation stage stepper motor is commanded to perform a ``scan", defined as a 0$^\circ$ $\rightarrow$ 360$^\circ$ rotation of the source at a specified step size and integration time per step. The integration time is determined by the desired signal-to-noise of the detector response which, for the calibrator coupled to the POLARBEAR telescope, was calculated to be five seconds to ensure a signal-to-noise of over 100. An MCCDAQ\footnote{Measurement Computing Corporation. Norton, MA. www.mccdaq.com} USB-1208FS data acquisition device (DAQ) is used to record the output of the two chopper encoders and a 360$^\circ$ analog tilt stage mounted on the rotation stage for redundant mid-measurement angle validation. The DAQ also monitors an enclosure-mounted encoder coupled to the rotation stage that is used to initialize the source polarization axis before performing a scan.

\section{DATA AND ANALYSIS}
\label{sec:analysis}

\subsection{Model}
\label{sec:model}

Detector response is continuously recorded as the calibrator source performs repeated scans. The detector timestream is separated according to the calibrator source angle and then demodulated to determine the power incident on the detector. The demodulated signals are fit to the model according to Eq.~\ref{eq:model} to recover the initial offset angle $\phi$ between the calibrator source and detector polarization planes. If the initial calibrator source polarization plane is referenced to the local gravity vector as described in Section~\ref{sec:calgrav}, then $\phi$ becomes the absolute detector polarization angle orientation with respect to the local gravity vector.

\begin{figure} [ht]
	\begin{center}
		\begin{tabular}{c} 
			\includegraphics[height=12cm]{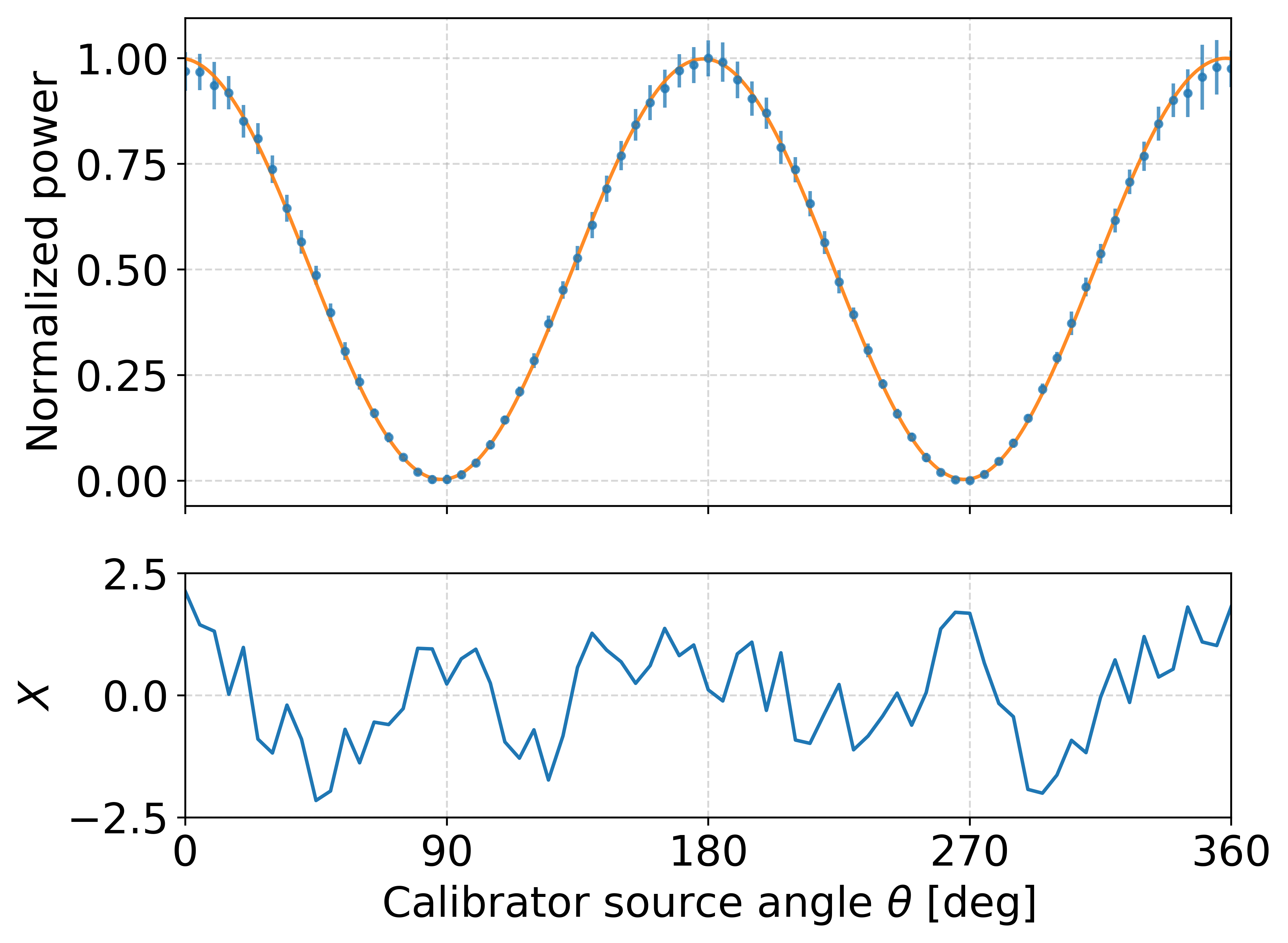} 
		\end{tabular}
	\end{center}
	\caption[example] 
	{ \label{fig:bolofit} 
		\textit{Top:} Normalized, demodulated laboratory detector power response (blue) for one calibrator scan fitted to the model in Eq.~\ref{eq:model} (orange). Error bars for detector power are larger when the source and detector are aligned due to systematic preamplifier noise. \textit{Bottom:} Residual between the data and the model normalized by the error at each source angle, with $X = (S_{fit} - S_{data}) / \sigma_{data}$.}
\end{figure}

The ideal demodulated detector signal $S(\theta)$ is expected to follow a cosine curve with a minimum of zero (as negative detector signal is non-physical) described by a model of the form
\begin{equation}
\label{eq:ideal}
S_{ideal}(\theta) = A\big(\cos\big(2(\theta - \phi)\big) + 1\big) + B \, ,
\end{equation}
where $A$ is the detector signal amplitude, $\theta$ is the source angle, $\phi$ is the initial angle offset between the source and the detector, and $B$ is a DC offset term. If the source is poorly collimated about its rotation axis, Eq.~\ref{eq:ideal} must be modified to account for the resulting precession and can be accounted for by adjusting the model as
\begin{equation}
\label{eq:model}
S(\theta) = \Big(\big(A\cos\big(2(\theta - \phi)\big) + 1\big) + B \Big) \Big(C \cos(\theta - \psi) + 1 \Big) \, ,
\end{equation}
where C and $\psi$ are the precession amplitude and phase. Fig.~\ref{fig:bolofit} shows an example demodulated detector power timestream from laboratory characterization (described later in Section~\ref{sec:labtests}).

\subsection{Systematics}

Table~\ref{tab:systematics} details relevant systematic uncertainties involved in a measurement of absolute detector angle polarization using the calibrator on the POLARBEAR telescope. The total calibrator polarization angle error is calculated to be $\pm0.025^\circ$ (statistical) and $\pm0.055^\circ$ (systematic). The statistical errors arise from bubble level measurements of the system while calibrating to the local gravity vector while the systematic errors are estimated by propagating the corresponding systematic model through the analysis pipeline. The systematic errors are estimated for deployment of the calibrator on the POLARBEAR telescope with several assumptions. The calibrator is assumed to be placed at a distance of about 400 m from the telescope, at which distance the FWHM of the calibrator beam projected on to the POLARBEAR focal plane is calculated to be about 40 mm in diameter. POLARBEAR bolometers are read out using frequency domain multiplexed superconducting quantum interference devices (SQUIDs), with eight bolometers read out per SQUID card. It is assumed that neighboring bolometers on any particular SQUID card experience electrical crosstalk of 2\%~\cite{Ade2014e} and that the two bolometers are spaced on the focal plane such that they are not both within the main lobe of the calibrator beam. Ground reflections were estimated to have a small effect on polarization angle calibration due to the extreme reflected angles being well outside the main lobe of the telescope beam~\cite{Ade2014e}. However, the reflection effect becomes much more pronounced and must be addressed at far-field calibrator distances. Calibrator beam deformities, Gunn oscillator temperature stability, and birefringence of the MF-110 attenuation slabs have the potential to generate large polarization angle errors but can be measured and accounted for to reduce errors to negligible size.
\begin{table}[ht]
	\caption{Calculated and estimated statistical and systematic errors.} 
	\label{tab:systematics}
	\begin{center}       
		\begin{tabular}{|l|l|} 
			\hline
			\rule[-1ex]{0pt}{3.5ex}  \textbf{Statistical uncertainties} & \textbf{Angle}  \\
			\hline
			\rule[-1ex]{0pt}{3.5ex}  Wire-grid wire wrapping & 0.02$^\circ$    \\
			\rule[-1ex]{0pt}{3.5ex}  Wire-grid misalignment & 0.006$^\circ$    \\
			\rule[-1ex]{0pt}{3.5ex}  Rotation stage backlash & 0.006$^\circ$    \\
			\rule[-1ex]{0pt}{3.5ex}  Pre-pointing gravity vector leveling & 0.006$^\circ$   \\
			\rule[-1ex]{0pt}{3.5ex}  Post-pointing & 0.006$^\circ$    \\
			\rule[-1ex]{0pt}{3.5ex}  \textbf{Quadrature sum} & \textbf{0.025$^\circ$}    \\
			\hline \hline
			\rule[-1ex]{0pt}{3.5ex}  \textbf{Systematic uncertainties} & \textbf{Angle}    \\
			\hline
			\rule[-1ex]{0pt}{3.5ex}  Electrical crosstalk & 0.05$^\circ$    \\
			\rule[-1ex]{0pt}{3.5ex}  Ground reflections & 0.015$^\circ$    \\
			\rule[-1ex]{0pt}{3.5ex}  Calibrator beam deformities & \textless0.01$^\circ$   \\
			\rule[-1ex]{0pt}{3.5ex}  Gunn diode temperature stability & \textless0.01$^\circ$   \\
			\rule[-1ex]{0pt}{3.5ex}  Birefringent MF-110 attenuators & \textless0.01$^\circ$   \\
			\rule[-1ex]{0pt}{3.5ex}  \textbf{Quadrature sum} & \textbf{0.055$^\circ$}    \\
			\hline
		\end{tabular}
	\end{center}
\end{table}

\subsection{Laboratory characterization}
\label{sec:labtests}

A room-temperature, broadband Pacific Millimeter~\footnote{Pacific Millimeter Products. Golden, CO. www.pacificmillimeter.com} 110-170 GHz detector diode coupled to a WR-06 pyramidal horn antenna, identical to the calibrator horn antenna, was used in conjunction with a low-noise voltage preamplifier to characterize the calibrator signal. The detector and horn antenna were mounted on a two-axis beam mapper consisting of two perpendicular linear stages coupled to separate stepper motors to precisely control the position of the detector relative to the calibrator. Eccosorb AN-72 was placed around both the calibrator and detector beams to mitigate reflections and ensure beam purity.

\begin{figure} [ht]
	\begin{center}
		\begin{tabular}{c} 
			\includegraphics[height=11cm]{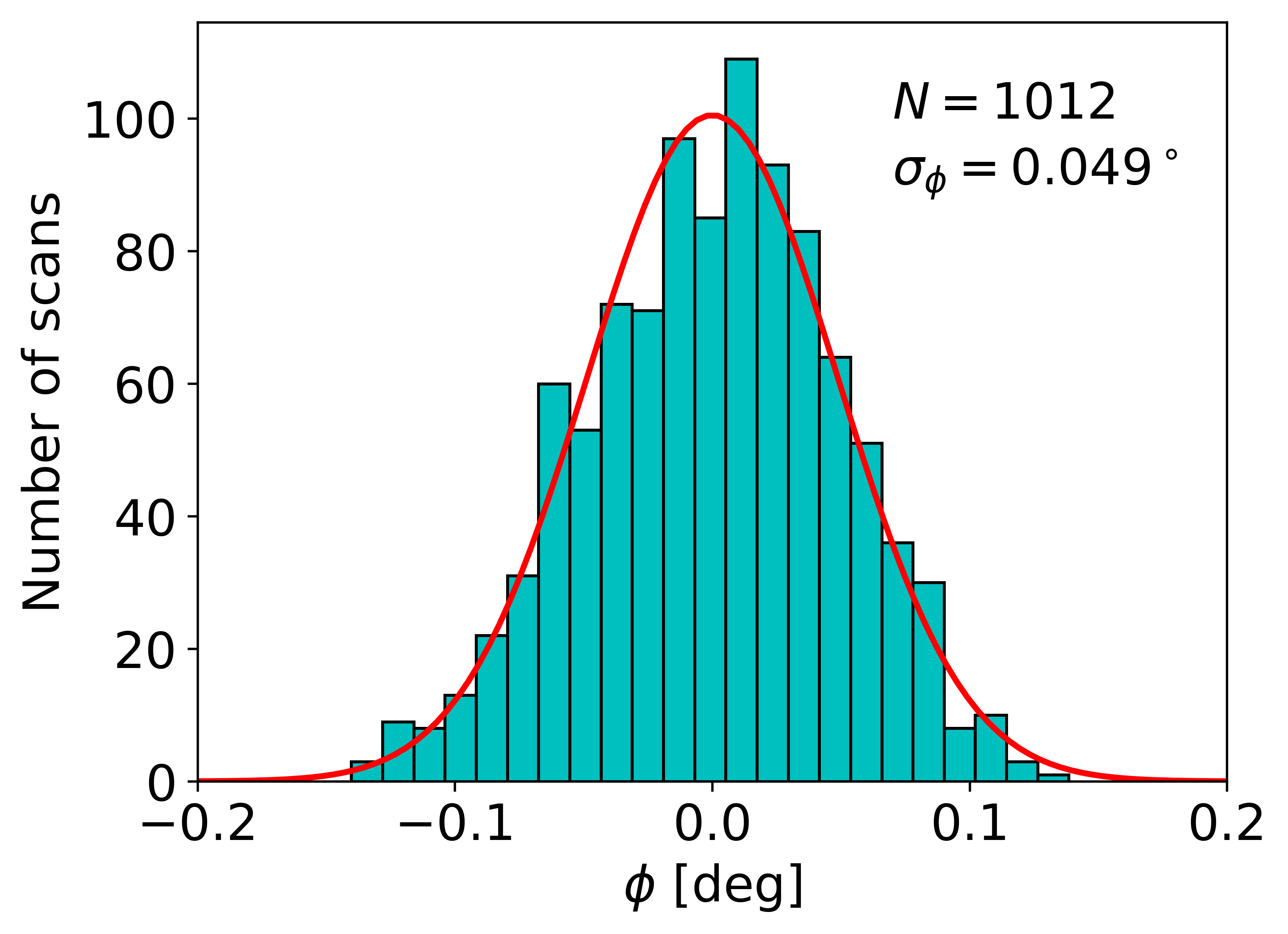}
		\end{tabular}
	\end{center}
	\caption[example] 
	{ \label{fig:hist} 
		Histogram of calibrator repeatability over N = 1012 scans. $\phi$ represents the misalignment angle between the initial source and detector polarization axes and was measured with an error of $\sigma_{\phi} = 0.049^\circ$ over the 1012 scans.}
\end{figure}

The consistency of the calibrator ability to measure the offset angle $\phi$ between the initial source and detector polarization planes was determined by fixing the position of the detector on the beam mapper and performing repeated 0$^\circ \rightarrow$ 360$^\circ$ scans of the source. Calibrator consistency tests were performed on six separate days with varying source and detector setups. After cutting scans with obvious glitches in their respective timestreams, the demodulated detector signal timestreams from 1012 scans taken over the six days were fitted according to the model given in Eq.\ref{eq:model}. The offset angle $\phi$ was measured to within $\pm 0.049^\circ$ over these 1012 scans (as shown in Fig.~\ref{fig:hist}) confirming the calibrator's repeatability to within the 0.1$^\circ$ goal.

\begin{figure} [ht]
	\begin{center}
		\begin{tabular}{c} 
			\includegraphics[height=6.65cm]{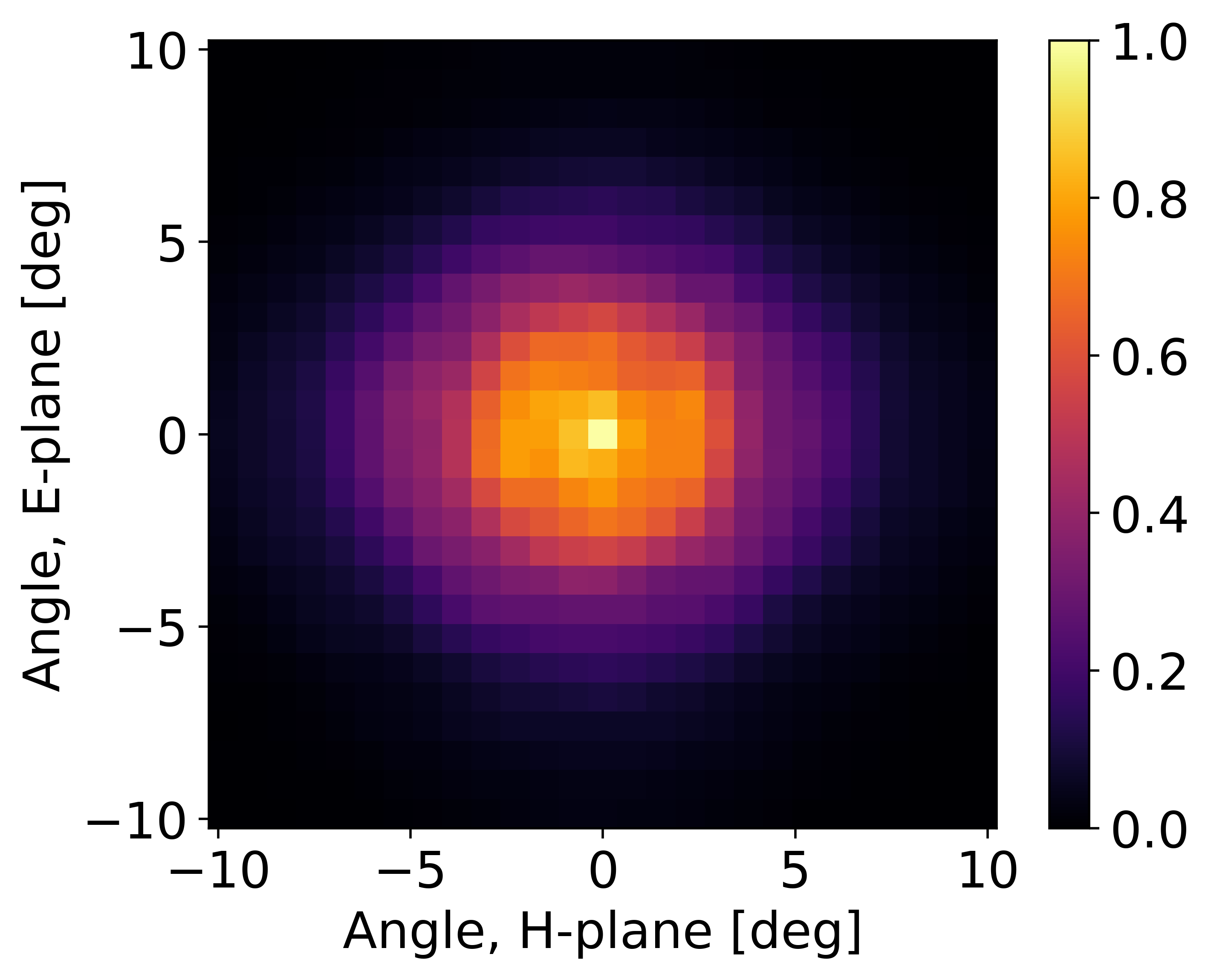}
			\includegraphics[height=6.6cm]{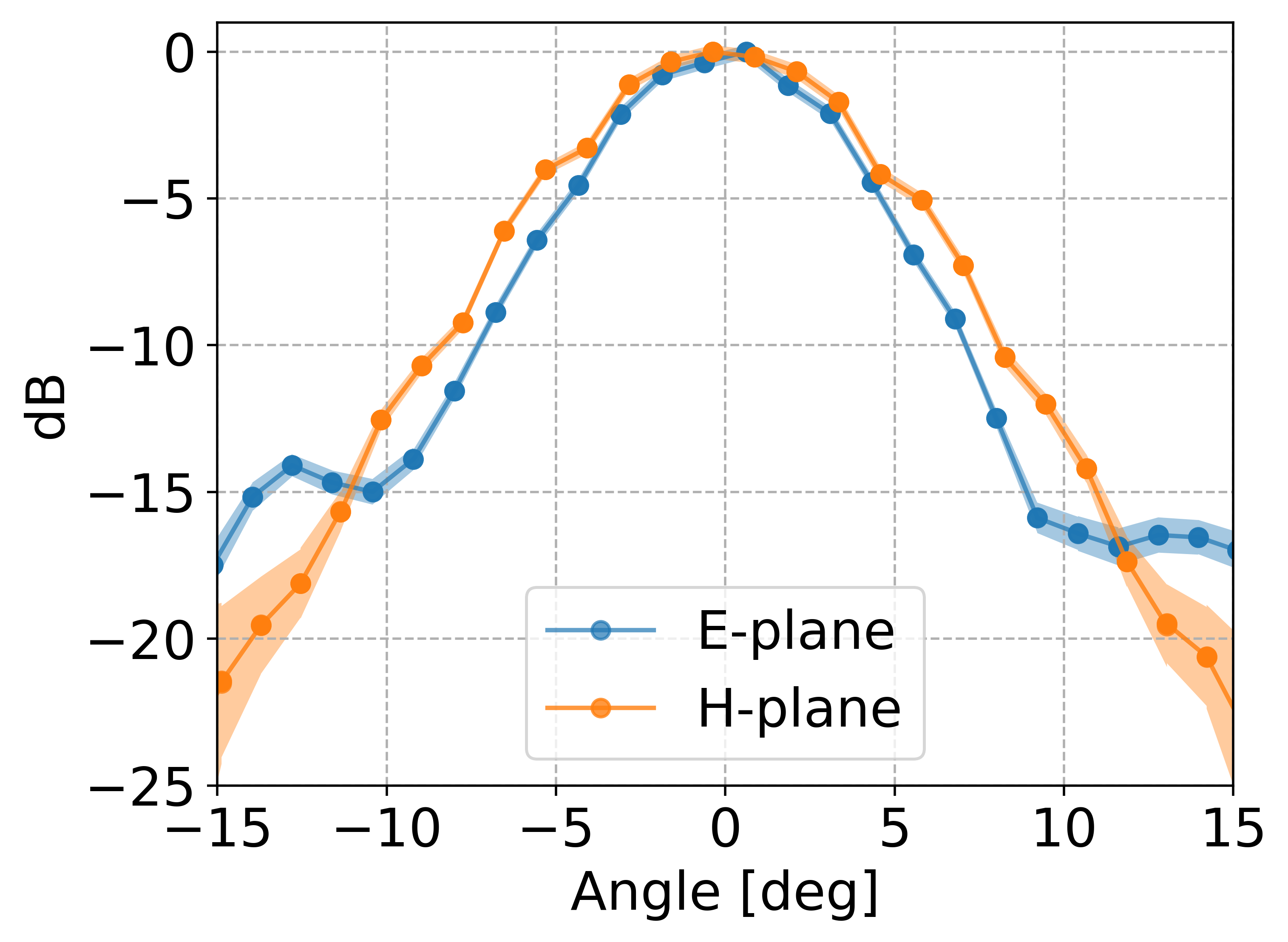}
		\end{tabular}
	\end{center}
	\caption[example] 
	{ \label{fig:beammap} 
		\textit{Left:} Beam map of the source and room-temperature detector diode. The two axes represent the E- and H-planes for the two horns and are plotted in degrees. The scale of the heat map is normalized to the maximum beam map value. \textit{Right:} Measured E- and H-plane beam profiles with errors given by the respective shaded regions.}
\end{figure}

Measurements of the calibrator beam were made by fixing the calibrator source angle and measuring the detector output at various positions using the beam mapper. The beam maps were produced by raster scanning 9~cm by 9~cm in 0.635~cm steps at a source distance from the detector of 47~cm (which translates to a 10.7$^\circ$ by 10.7$^\circ$ map in 0.77$^\circ$ steps) and are shown in Fig.~\ref{fig:beammap}. All beam map measurements were taken in the far-field of both horn antennas (each of which have a far-field distance of 22.5~cm) and are fitted to a two-dimensional Gaussian function. From beam maps made at calibrator source angles of 0$^\circ$ and 180$^\circ$, the fitted beam center for the two angles revealed a 0.5$^\circ$ source horn antenna axis misalignment with the calibrator rotation axis as described in Section~\ref{sec:model}. Fig.~\ref{fig:beammap} also shows the measured E- and H-plane beam profiles in units of dB.

\subsection{Deployment on the POLARBEAR telescope}
\label{sec:deployment}

The calibrator was deployed for an engineering test run on the POLARBEAR telescope in March 2017. The primary goals of the deployment were to prove coupling between the calibrator and detectors, validate calibrator control operation in the field, validate detector angle calibration analysis methodology in the field, and identify potential improvements to be implemented before a second calibrator deployment. The March 2017 deployment was successful in achieving each of these goals. The chopped calibrator signal was clearly visible in the POLARBEAR detector timestreams, the calibrator control software correctly executed over the course of several days, the calibrator analysis pipeline successfully demodulated timestreams from POLARBEAR detectors with expected sinusoidal detector response over calibrator scans, and several potential improvements (discussed in Section~\ref{sec:future}) were noted throughout the course of the deployment.

Attempts to measure the relative and absolute detector polarization angle orientations were made after successful validation of the deployment's primary goals. However, the detector calibration angles measured were not more accurate than POLARBEAR's existing detector angle calibration methods due to unexpected issues with telescope control hardware, detector saturation, and calibrator chopper motor operation. However, the deployment achieved its primary goals and the calibrator will be improved to more accurately measure absolute detector polarization angles for future CMB experiments at the site.

\section{FUTURE APPLICATION}
\label{sec:future}

Future CMB experiments such as the Simons Array and the Simons Observatory will observe in several frequency bands~\cite{Suzuki2016,Stebor2016}. To meet the demands to calibrate the absolute detector polarization orientation for these experiments in all possible observing bands, the current calibrator will be upgraded to contain multiple tunable RF sources at the appropriate band center frequencies, specifically at 40/95/150/220 $\pm10\%$ GHz. To achieve the desired bandwidth for each band, the Gunn oscillator will be replaced with either a tunable Gunn diode oscillator or phase locked synthesizer (PLS). The calibrator will be designed to easily accommodate extra sources in the 30 and 270 GHz bands. From the engineering test deployment described in section~\ref{sec:deployment}, we determine a number of other upgrades to the current calibrator that will be made, summarized in Table~\ref{tab:upgrades}. 

\begin{table}[ht]
	\caption{Summary of forthcoming calibrator upgrades compatible with future CMB experiments.} 
	\label{tab:upgrades}
	\begin{center}       
		\begin{tabular}{|l|l|l|} 
			\hline
			\rule[-1ex]{0pt}{3.5ex}  \textbf{Component} & \textbf{Current Calibrator} & \textbf{Upgraded calibrator}  \\
			\hline
			\rule[-1ex]{0pt}{3.5ex}  Source & Gunn oscillator & Tunable Gunn oscillator or PLS   \\
			\hline
			\rule[-1ex]{0pt}{3.5ex}  Band Center Frequency & 150 GHz & 40/95/150/220 GHz    \\
			\hline
			\rule[-1ex]{0pt}{3.5ex}  Tunable Bandwidth & N/A & $\pm 10\%$    \\
			\hline
			\rule[-1ex]{0pt}{3.5ex}  Chopper & Physical wheel & PIN switch  \\
			\hline
			\rule[-1ex]{0pt}{3.5ex}  Horn/rotation axis misalignment  & N/A & Tip-tilt stage source mount   \\
			\hline
			\rule[-1ex]{0pt}{3.5ex}  Attenuation & MF-110 slabs & Level set attenuators \\
			\hline
			\rule[-1ex]{0pt}{3.5ex}  Gravity vector calibration & Two-axis tilt stage & Multiple digital tiltmeters   \\
			\hline
		\end{tabular}
	\end{center}
\end{table}

The calibrator's physical chopper wheel will be replaced by a PIN switch in the source circuit. A PIN switch will be superior to the physical chopper wheel as it will be able to modulate the calibrator signal at a much higher rate, will simplify the control and analysis procedure by removing the dependence on the two physical chopper encoders, and will remove excess vibrations generated by the current chopper motor. As described in Section~\ref{sec:labtests}, the current source horn antenna axis is misaligned with the calibrator rotation axis by up to 0.5$^\circ$ with no option for adjustment. The tunable RF source will be mounted on an adjustable tip-tilt stage for fine control of the source horn antenna axis. The calibrator's dependence on physical MF-110 attenuator sheets, which could rotate the source polarization via stress induced birefringence from machining or distort the beam via reflections, will be removed and replaced with level set attenuators in the source circuit. The upgraded RF source circuit can be summarized by the following chain: tunable RF source $\rightarrow$ multiplier $\rightarrow$ PIN switch (driven by an input TTL signal) $\rightarrow$ level set attenuators $\rightarrow$ pyramidal horn antenna. The current procedure to calibrate the source polarization angle to the local gravity vector as described in section~\ref{sec:calgrav} currently depends heavily on leveling the two-axis tilt stage (plane D in Fig,~\ref{fig:planes}) twice and is extremely time consuming. This leveling procedure will be significantly simplified by adding continuously sampled, high sensitivity digital tiltmeters on planes A, B, and C in Fig.~\ref{fig:planes}, which were not included in the initial calibrator design.

\section{CONCLUSIONS}
\label{sec:conclusions}

The ground-based absolute detector polarization angle calibrator for use with CMB experiments discussed in this paper utilizes a rotating polarized microwave source with co-rotating polarizing wire grid and was designed to be compact, robust to harsh environments, and accurate to better than 0.1$^\circ$ in order to reduce upper limits on the CMB TB and EB correlations and allow for the search of evidence of CPR effects such as cosmic birefringence and PMFs. The calibrator was proven in the laboratory to a consistency within the 0.1$^\circ$ requirement. From its deployment in March 2017 on the POLARBEAR telescope, the calibrator was proven to successfully couple to POLARBEAR detectors, the control and analysis procedure was verified, and detector response was observed to exhibit the expected behavior when rotating the calibrator source. The successful laboratory characterization combined with the successful coupling of the calibrator to the POLARBEAR telescope proves that this type of calibrator design can reach the 0.1$^\circ$ accuracy requirement and is being improved to perform calibration of future CMB experiments. The improved calibrator design is expected to be built and characterized in the laboratory by the end of 2018 using the POLARBEAR-2b receiver and will be compatible with future CMB experiments such as the Simons Array and the Simons Observatory.

\acknowledgments 

POLARBEAR and The Simons Array is funded by the Simons Foundation and by grants from the National Science Foundation AST-0618398 and AST-1212230. The Simons Array will operate at the James Ax Observatory in the Parque Astronomico Atacama in Northern Chile under the stewardship of the Comisi\'{o}n Nacional de Investigacion Cientif\'{i}ca y Tecnol\'{o}gica de Chile (CONICYT). The polarizing wire grid and the beam mapper used in the characterization of the calibrator were both designed and constructed at UC San Diego. The power and frequency output of the Gunn oscillator were measured at UC Berkeley. The attenuation of MF-110 was measured at the High Energy Accelerator Research Organization (KEK) in Tsukuba, Ibaraki 305-0801, Japan.

\bibliography{bibtex/polcal.bib} 
\bibliographystyle{spiebib} 

\end{document}